\documentclass[twocolumn,trackchanges]{aastex701}

	\newcommand{\bqn}{\begin{eqnarray}}
	\newcommand{\eqn}{\end{eqnarray}}
	\newcommand{\nb}{\nonumber}
	
	\newcommand{\f}{\frac}
	\newcommand{\p}{\partial}

    \usepackage{graphicx}	
    \usepackage{amsmath, mathrsfs}	
    \usepackage{amssymb}	

\begin{document}

\title{Relativistic Thermal Emission from Accretion Disks in Kerr-MOG Spacetimes}

\author[0000-0002-2714-9257]{Cheng Liu}
\affiliation{Tsung-Dao Lee Institute, Shanghai Jiao Tong University, 1 Lisuo Road, Shanghai, 201210, China}
\email{liuc09@sjtu.edu.cn}

\author[0009-0006-9884-6128]{Xufan Hu}
\affiliation{Tsung-Dao Lee Institute, Shanghai Jiao Tong University, 1 Lisuo Road, Shanghai, 201210, China}
\email{ebr105@sjtu.edu.cn}

\author[0000-0002-8131-6730]{Yosuke Mizuno}
\email[show]{mizuno@sjtu.edu.cn}
\affiliation{Tsung-Dao Lee Institute, Shanghai Jiao Tong University, 1 Lisuo Road, Shanghai, 201210, China}
\affiliation{School of Physics and Astronomy, Shanghai Jiao Tong University, 
800 Dongchuan Road, Shanghai, 200240, China}
\affiliation{Key Laboratory for Particle Astrophysics and Cosmology (MOE) and Shanghai Key Laboratory for Particle Physics and Cosmology,
Shanghai Jiao Tong University, Shanghai 200240, China}
\affiliation{Institut f\"ur Theoretische Physik, Goethe-Universit\"at Frankfurt, Max-von-Laue-Stra{\ss}e 1, D-60438 Frankfurt am Main, Germany}
\author[0000-0003-2286-9009]{Tao Zhu}
\email[show]{zhut05@zjut.edu.cn}
\affiliation{Institute for Theoretical Physics \& Cosmology, Zhejiang University of Technology, Hangzhou, 310023, China}

\begin{abstract}
In Scalar-Tensor-Vector Gravity (STVG, also known as MOG), a massive vector field $\phi_\mu$ generates a repulsive fifth force that endows rotating black holes with a gravitational charge $Q \propto \sqrt{\alpha}\,M$, modifying the near-horizon geometry through a single deformation parameter $\alpha$. We investigate how this vector-field coupling imprints itself on the thermal continuum emission of geometrically thin, optically thick accretion disks in the Kerr-MOG black hole. By re-deriving the innermost stable circular orbit (ISCO), the Novikov-Thorne radiative flux, the relativistic energy shift, and the null geodesic structure for the Kerr-MOG spacetime, we compute fully relativistic disk spectra across a broad range of spins, inclinations, and fifth-force strengths using a dedicated \textsc{xspec} spectral model (\texttt{kmspec}). We find that the fifth-force charge pushes the ISCO outward, lowers the peak disk temperature, and systematically softens the thermal continuum relative to its Kerr black hole counterpart at the same spin, with the deviation amplified at high observer inclinations. The resulting spectral modification closely mimics a reduction of spin in the pure Kerr black hole framework, indicating that independent spin measurements from, e.g., iron-line reflection spectroscopy are indispensable for disentangling the vector-field contribution. All results recover the standard Kerr black hole predictions when $\alpha = 0$, and the model is validated against independent analytic and
numerical benchmarks to machine precision. Application to a 69.6~ks \textit{XMM-Newton} observation of LMC~X-1 yields $\alpha < 0.044$ at 90\% confidence, consistent with the Kerr metric and general relativity.
\end{abstract}


\section{Introduction} \label{sec:intro}

Tests on the theory of general relativity (GR) have been implemented since it was discovered in 1915. It is still the most accepted theory of gravity. 
Recently, the horizon-scale observation of supermassive black holes (SMBH) M\,87$^*$ and Sgr\,A$^*$ by Event Horizon Telescope Collaboration unveiled the physics of the black holes (BH) and the accretion flow surrounding them with the highest angular resolution that we can achieve on Earth \citep{EventHorizonTelescope:2019dse, Michail:2024zdp}. Although we still cannot distinguish Kerr BH and BH solutions from other theories of gravity \citep[e.g.,][]{Mizuno:2018lxz, 2020PhRvL.125n1104P,EHT2022f,Ozel2022,Afrin2023,2023arXiv231204288J,Younsi2023,2023CQGra..40p5007V, Uniyal:2025uvc}, an upper limit for the ``charge" of the BH can be measured \citep{EventHorizonTelescope:2021dqv, Kocherlakota:2022jnz,EHT2022f}. Within the acceptable parameter space for the SMBHs, there are still many unique features from alternative theories of gravity.

X-ray spectroscopy of accreting stellar-mass black holes offers one of the most direct probes of near-horizon spacetime geometry. The thermal continuum emitted by a geometrically thin, optically thick accretion disk encodes the location of the innermost stable circular orbit (ISCO), the radial temperature profile, and the relativistic energy shift between the emitting and observing frames \cite[e.g.,][]{Zhang1997, Steiner2010, McClintock2014}. Because these quantities depend sensitively on the underlying spacetime, continuum fitting has been developed into a precision tool for measuring black hole spin \citep{McClintock:2006xd, Gou2009, Steiner2010} and, more recently, for testing the Kerr hypothesis itself \citep{Bambi:2013sha, Zhou:2019fcg}. Application of this technique to non-Kerr spacetimes has yielded competitive constraints on deformation parameters in several frameworks \citep{Tripathi:2020qco, Zhang:2021ymo, Tripathi:2021rqs, Yu:2021xen}, establishing broadband X-ray observations as a complementary channel to gravitational-wave and shadow-based tests of strong-field gravity.

Scalar--Tensor--Vector Gravity (STVG/MOG; \citealt{Moffat:2005si}) augments Einstein's field equations with a Proca-type massive vector field $\phi_\mu$ that couples universally to baryonic stress-energy. This coupling is controlled by a single dimensionless parameter $\alpha$, which simultaneously enhances the effective gravitational constant and introduces a repulsive Yukawa-type correction at short range---a ``fifth force'' \citep{Moffat:2015kva, Green:2017qcv}. The theory was originally constructed to reproduce galactic rotation curves without invoking cold dark matter \citep{Bertone:2004pz, Brownstein:2007sr, Moffat:2013sja}. Its strong-field sector admits the Kerr-MOG rotating black hole solution \citep{Moffat:2015kva}, whose geometry departs from Kerr black hole through a single charge-like contribution proportional to~$\alpha$.

The astrophysical implications of this deformation in MOG have been extensively explored across several domains. Photon orbits and shadow morphology were investigated by \citep{Moffat:2015kva, Wang:2018prk, Kuang:2022ojj, Qin:2022kaf, Zhang:2024jrw, Zheng:2024ftk}, while quasi-normal modes and gravitational wave constraints were addressed in \citep{Manfredi:2017tdn} and \citep{Moffat:2016gkd}, respectively. Beyond horizon thermodynamics \citep{Mureika:2015sda}, significant focus has been placed on particle dynamics, specifically timelike geodesics and the innermost stable circular orbit (ISCO) \citep{Hussain:2015cga, Lee:2017fbq, Sharif:2017owq, Sheoran:2017dwb, Wei:2018aft}. Furthermore, the structure and observable signatures of accretion disks have been studied through both theoretical modeling \citep{Perez:2017spz} and high-resolution X-ray spectroscopy \citep{Miller:2025aqd}. However, the observable most directly tied to the ISCO structure, the thermal X-ray continuum, has not been treated in a form suitable for quantitative spectral fitting. Although dedicated \textsc{xspec} models now exist for several non-Kerr spacetime \citep{Bambi:2016sac, Bambi:2015kza, Tripathi:2020yts, Zhou:2019fcg}. However, no analogous spectral package has been constructed for the Kerr-MOG black hole, leaving the theory without a self-consistent pathway to confront broadband X-ray data of accreting black holes.

In this paper, we address this deficiency by conducting a systematic study of how the fifth-force vector charge reshapes the thermal continuum of geometrically-thin accretion disks in the Kerr-MOG spacetime. We develop \texttt{kmspec}, a fully relativistic spectral model implemented as a local \textsc{xspec} package, which re-derives the disk physics and geodesic equations for the MOG-modified metric by mapping them onto a relativistic ray-tracing framework. The primary objectives of this work are to quantify the spectral imprints of the vector-field coupling across the $(a_*,\,\alpha)$ parameter space and to evaluate the degree to which these imprints are degenerate with the black hole spin. This analysis provides a robust pathway to test the viability of MOG against high-quality X-ray observations of black hole candidates.

The remainder of the paper is organized as follows. Section~\ref{sec:theory} reviews the MOG vector field as a fifth force and presents the Kerr-MOG metric. Section~\ref{sec:mods} details the physical ingredients of the spectral model construction. Section~\ref{sec:results} presents verification tests and representative spectral results. Section~\ref{sec:xrism} presents simulated \textit{XRISM} and \textit{XMM-Newton} observations and quantifies the detectability of the MOG signature. Section~\ref{sec:constraints} derives constraints on $\alpha$ from spectral fitting. Section~\ref{sec:concl} collects the conclusions.

\section{STVG and Kerr-MOG Solution}\label{sec:theory}

\subsection{MOG as a Fifth Force}\label{sec:5thforce}

The Scalar–Tensor–Vector Gravity (STVG) theory supplements the Einstein–Hilbert action with a massive vector field $\phi_\mu$, whose field strength $B_{\mu\nu} = \partial_\mu \phi_\nu - \partial_\nu \phi_\mu$ contributes a kinetic term $1/4 B_{\mu\nu}B^{\mu\nu}$ and a mass term $1/2 \mu^2 \phi_\mu \phi^\mu$. This vector field mediates a Yukawa-type fifth force between baryonic sources \citep{Moffat:2005si}. The coupling strength is governed by a dimensionless parameter $\alpha$, which defines an effective gravitational charge
\bqn
Q_g = \sqrt{\alpha G_N} M.
\eqn
In the strong-field regime, the running couplings approach constant background values, such that the theory reduces to an effective Einstein–Proca system with $G_{\mathrm{eff}} = G_N(1+\alpha),\, \mu = \mathrm{const.}$
In this limit, the theory admits a stationary, axisymmetric black-hole solution known as the Kerr–MOG metric \citep{Moffat:2015kva}.

The gravitational charge $Q_g$ is not of electromagnetic origin; rather, it parametrizes the repulsive contribution of the vector field at distances $r \lesssim \lambda_\phi = \mu^{-1}$, while being exponentially suppressed at larger radii. This additional degree of freedom modifies the effective gravitational interaction and provides a key mechanism through which STVG departs from GR in strong-field environments.

\subsection{Kerr--MOG Spacetime} \label{sec:metric}

An exact rotating black-hole solution in STVG is given by the Kerr–MOG spacetime \citep{Moffat:2015kva}, which in Boyer–Lindquist coordinates $(t,r,\theta,\varphi)$ takes the form
\begin{align}    
ds^2 = & -\frac{\Delta - a^2\sin^2\theta}{\rho^2}dt^2      - \frac{2a\sin^2\theta\,(r^2 + a^2 - \Delta)}{\rho^2}\,dt\,d\varphi \nonumber\\     & + \frac{\rho^2}{\Delta}dr^2 + \rho^2 d\theta^2      +  \frac{(r^2 + a^2)^2 - a^2\Delta \sin^2\theta}{\rho^2} \sin^2\theta d\varphi^2, 
\end{align}
where
\bqn
\rho^2 = r^2 + a^2 \cos^2\theta,
\eqn
and
\bqn
\Delta = r^2 - 2G_N(1+\alpha)Mr + a^2 + \alpha(1+\alpha)G_N^2 M^2.
\eqn
This metric is formally analogous to the Kerr–Newman solution, with the square of the electric charge replaced by an effective gravitational charge
\bqn
Q_{\rm eff}^2 = \alpha(1+\alpha)G_N^2 M^2 = (1+\alpha) Q_g^2,
\eqn
which makes explicit the relation between the metric charge and the fundamental gravitational charge introduced above.

The spacetime is characterized by the mass $M$ and the spin parameter $a = J/M$, where $J$ is the angular momentum. We define the gravitational radius $r_g \equiv G_N M / c^2$ and the dimensionless spin parameter $a_* = a/r_g$.
In geometric units ($G_N = c = M = 1$), one has $r_g = 1$ and $a_* = a$. The Kerr solution is recovered in the limit $\alpha \to 0$.

The event horizons are determined by the roots of $\Delta = 0$, yielding
\bqn
r_{\pm} = G_N(1+\alpha)M \pm \sqrt{G_N^2(1+\alpha)M^2 - a^2},
\eqn
where the contribution of the effective charge has been absorbed into the modified mass term. In geometric units, the metric function simplifies to
\bqn
\Delta = r^2 - 2(1+\alpha)r + a^2 + \alpha(1+\alpha),
\eqn
making explicit that deviations from the Kerr geometry arise from both an enhanced effective gravitational coupling and a repulsive charge-like term. 

For $\alpha > 0$, these modifications alter the structure of timelike and null geodesics, thereby affecting both particle dynamics and photon propagation in the vicinity of the black hole. This provides the foundation for computing observable signatures, such as the thermal emission from accretion disks, which we develop in the following section.

\section{Spectral Model Construction}\label{sec:mods}

To consistently incorporate the effects of the MOG fifth force into the thermal continuum emission, we construct a Kerr--MOG spectral model by extending a Kerr--Newman framework. All quantities entering the disk emission and photon propagation are generalized by replacing the electromagnetic charge parameter with the MOG deformation parameter $\alpha$, ensuring that its influence is propagated self-consistently through the spacetime geometry, particle orbits, and radiative transfer.

\subsection{Disk Structure and Emission}\label{sec:disk}

The accretion disk is modeled within the Novikov--Thorne thin-disk framework \citep{NovikovThorne1973,Page1974}. The inner edge of the disk is set by the innermost stable circular orbit (ISCO), defined as the marginally stable circular geodesic in the equatorial plane.  For a massive test particle on an equatorial orbit, the radial equation of motion can be cast in the form $\dot{r}^{2} = -V_{\mathrm{eff}}(r;\,E,\,L)$, where $E$ and $L$ are the specific energy and angular momentum, and $V_{\mathrm{eff}}$ is the effective potential.  Circular orbits require $V_{\mathrm{eff}} = 0$ and $V_{\mathrm{eff}}' = 0$, which fix the orbital energy and angular momentum as functions of $r$.  Stability demands $V_{\mathrm{eff}}'' \geq 0$; the ISCO is the orbit at which equality holds,
\bqn
  V_{\mathrm{eff}} = 0,\qquad
  V_{\mathrm{eff}}' = 0,\qquad
  V_{\mathrm{eff}}'' = 0.
\eqn
This triple condition is equivalent to requiring $dE_{\mathrm{circ}}/dr = 0$ (or, equivalently, $dL_{\mathrm{circ}}/dr = 0$), i.e., the ISCO is the radius at which the specific orbital energy of circular geodesics reaches its minimum \citep{Lee:2017fbq}.

For $\alpha = 0$, the ISCO reduces to the analytic expression for the Kerr black hole \citet{Bardeen1972}.  For $\alpha > 0$, the resulting equation involves fractional powers, and no closed-form solution exists in the rotating case \citep{Lee:2017fbq}; the ISCO is therefore located numerically. The specific orbital energy of circular geodesics is
\bqn
  E_{\mathrm{circ}}(r) =
    -\frac{g_{tt} + \Omega_K\,g_{t\varphi}}
          {\sqrt{-g_{tt} - 2\Omega_K g_{t\varphi} - \Omega_K^{2}\,g_{\varphi\varphi}}}\,,
\eqn
where the Keplerian angular velocity is
\bqn
  \Omega_K = \frac{\sqrt{M_{\mathrm{eff}}\,r - Q_{\mathrm{eff}}^{2}}}
                  {r^{2} + a\sqrt{M_{\mathrm{eff}}\,r - Q_{\mathrm{eff}}^{2}}}\,.
\eqn
The corresponding specific angular momentum of a circular orbit is
\bqn
  L_{\mathrm{circ}}(r) =
    \frac{g_{t\varphi} + \Omega_K\,g_{\varphi\varphi}}
         {\sqrt{-g_{tt} - 2\Omega_K g_{t\varphi} - \Omega_K^{2}\,g_{\varphi\varphi}}}\,.
\eqn

\begin{figure}[t]
  \centering
    \includegraphics[width=\columnwidth]{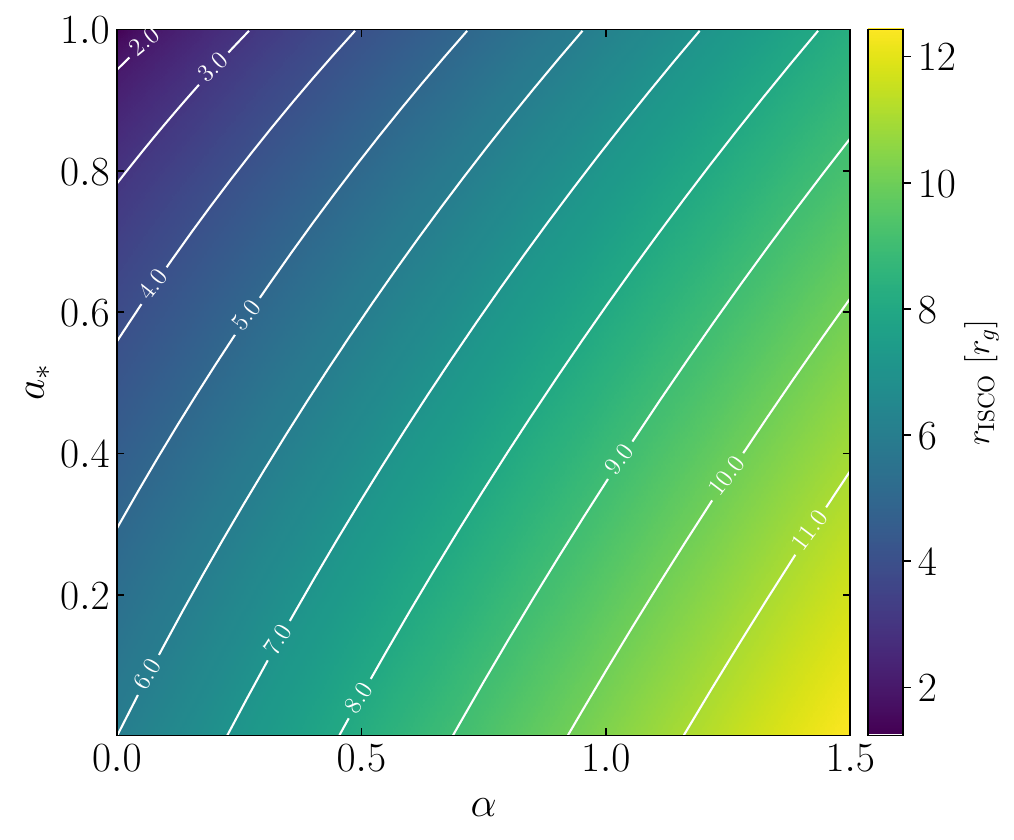}
  \caption{%
    ISCO radius $r_{\mathrm{ISCO}}$ (color scale, in units of $r_g$) as a function of $\alpha$ and spin $a_*$.  White contours are labeled in $r_g$.  
    At $\alpha = 0$, the Kerr black hole values are recovered. The monotonic outward shift with increasing $\alpha$ is evident at all spins.  Contours of constant $r_{\mathrm{ISCO}}$ run approximately diagonally, illustrating the $(a_*,\alpha)$ degeneracy discussed in Section~\ref{sec:interp}.}
  \label{fig:isco_alpha}
\end{figure}

\begin{figure}[t]
  \centering
    \includegraphics[width=\columnwidth]{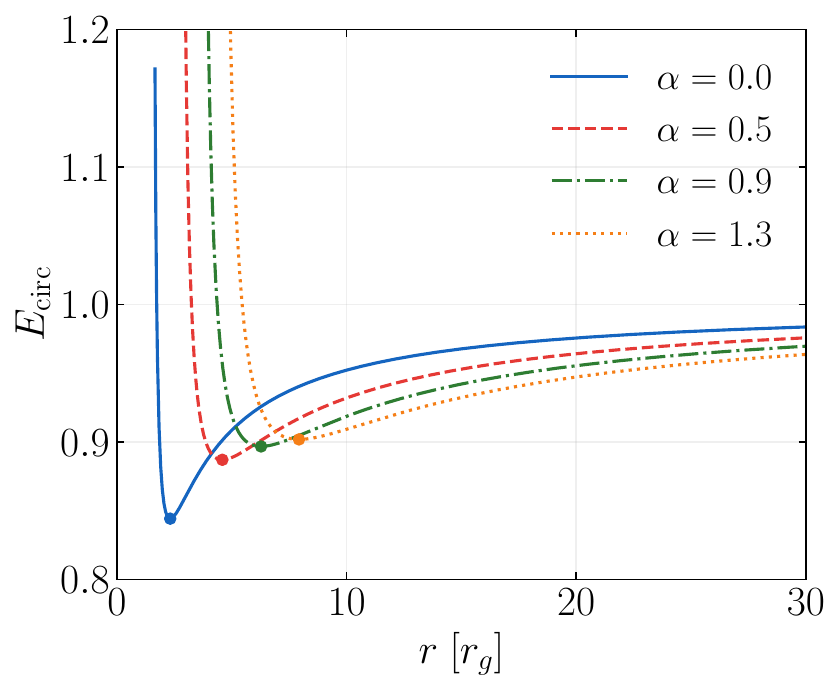}
  \caption{%
    Circular-orbit specific energy $E_{\mathrm{circ}}(r)$ for
    $\alpha = 0,\,0.5,\,0.9,\,1.3$ at $a_* = 0.9$.
    Filled circles mark the ISCO (energy minimum).
    Increasing $\alpha$ shifts the ISCO outward and
    reduces the binding energy at the marginally stable orbit.}
  \label{fig:effective_potential}
\end{figure}

Figure~\ref{fig:isco_alpha} presents the ISCO radius across the full $(\alpha,\,a_*)$ parameter space. At $\alpha = 0$, the standard Kerr black hole values are recovered, ranging from $r_{\mathrm{ISCO}} = 6\,r_g$ for a Schwarzschild black hole ($a_* = 0$) down to $r_{\mathrm{ISCO}} \simeq 1.24\,r_g$ at $a_* = 0.998$.  Activating the fifth force pushes the ISCO outward monotonically at every spin: the contours of constant $r_{\mathrm{ISCO}}$ run approximately diagonally from high spin/low $\alpha$ to low spin/high $\alpha$, demonstrating that the mapping $r_{\mathrm{ISCO}}(a_*,\alpha)$ admits a family of $(a_*,\alpha)$ pairs that produce the same ISCO location. This degeneracy is the geometrical origin of the spin--fifth-force confusion discussed in Section~\ref{sec:interp}. For slowly spinning black holes, the contours are nearly vertical (the ISCO depends weakly on $\alpha$), whereas for rapidly spinning black holes, even modest $\alpha$ induces a substantial outward shift because the ISCO of a Kerr black hole is already close to the horizon.  Figure~\ref{fig:effective_potential} visualizes this effect through the radial profile of $E_{\mathrm{circ}}(r)$ at $a_* = 0.9$: the energy minimum that defines the ISCO shifts to progressively larger radii with increasing $\alpha$, while the binding energy at the ISCO decreases.

In the standard Novikov--Thorne framework, the viscous torque at the ISCO is assumed to vanish (the ``vanishing ISCO stress'' or ``no-torque'' inner boundary condition; \citealt{NovikovThorne1973,Page1974}). The disk is assumed to be geometrically thin, optically thick, in a steady state, and confined to the equatorial plane; the mass accretion rate $\dot{M}$ is constant throughout the disk. The radiative flux emitted per unit proper area is then
\bqn
  \mathcal{F}(r) =
    -\frac{\Omega_K'}{4\pi\,r\,(E - \Omega_K L)^{2}}
    \int_{r_{\mathrm{ISCO}}}^{r}(E - \Omega_K L)\,
      \frac{dL}{dr'}\,dr'\,, \nb\\
\eqn
where a prime denotes a radial derivative, $E = E_{\mathrm{circ}}$ and $L = L_{\mathrm{circ}}$ are evaluated on circular orbits, and all quantities are computed self-consistently from the Kerr--MOG metric.  Relaxing this boundary condition introduces the dimensionless ISCO stress parameter $\delta_{\mathcal{J}} \in [0,\,1]$, which interpolates between vanishing stress at the ISCO ($\delta_{\mathcal{J}} = 0$, the standard assumption) and a finite residual torque ($\delta_{\mathcal{J}} > 0$). In this work, we set $\delta_{\mathcal{J}} = 0$ throughout; exploring non-zero values is deferred to future work. Here $\mathcal{F}(r)$ is a dimensionless function that encodes the radial profile of the flux. The physical radiative flux per unit proper area is $\dot{M}\,\mathcal{F}(r)/(4\pi M^{2})$. The local effective temperature follows from the Stefan--Boltzmann law:
\bqn
  T(r) =
    \left[\frac{\dot{M}\,\mathcal{F}(r)}{4\pi\,\sigma\,M^{2}}\right]^{1/4}.
\eqn

The accretion rate is conveniently expressed in units of the Eddington rate. The Eddington luminosity is the critical luminosity at which radiation pressure on a fully ionized hydrogen plasma balances gravitational attraction,
\bqn
  L_{\mathrm{Edd}} = \frac{4\pi\,G M m_p c}{\sigma_T}
    \approx 1.26 \times 10^{31}\,
    \left(\frac{M}{M_\odot}\right)\ \mathrm{W},
\eqn
where $m_p$ is the proton mass and $\sigma_T$ is the Thomson cross section. The corresponding Eddington mass accretion rate is defined as $\dot{M}_{\mathrm{Edd}} \equiv L_{\mathrm{Edd}}/c^{2}$. Not all accreted rest-mass energy is radiated: the radiative efficiency
\bqn
  \eta = 1 - E_{\mathrm{circ}}(r_{\mathrm{ISCO}})
\eqn
gives the fraction of $\dot{M}\,c^{2}$ that is released as electromagnetic radiation. For a Kerr black hole, $\eta$ ranges from $\simeq 0.057$ (Schwarzschild) to $\simeq 0.42$ (maximal prograde spin). In the Kerr-MOG spacetime, $\eta$ depends on $\alpha$ as well because the outward shift of the ISCO reduces the binding energy at its location.

\begin{figure*}[t]
  \centering
    \includegraphics[width=\textwidth]{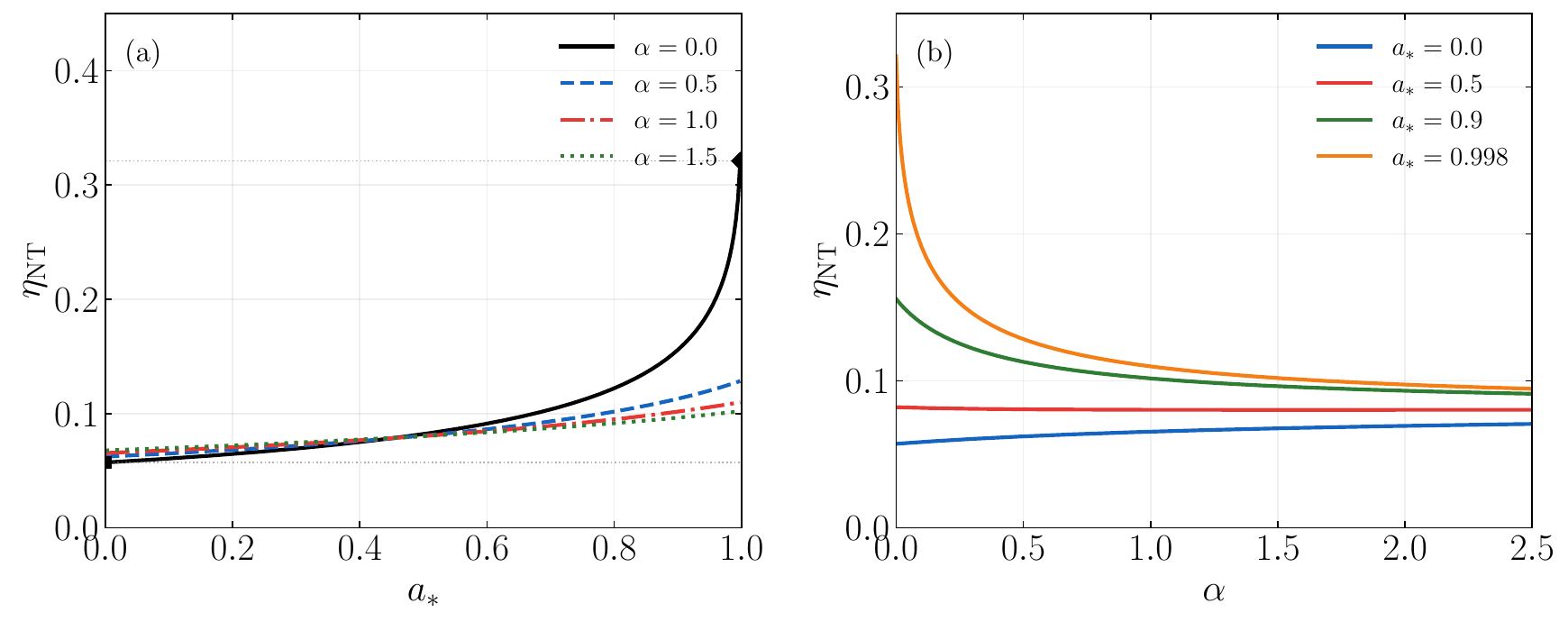}%
  \caption{%
    Novikov--Thorne radiative efficiency $\eta = 1 - E_{\mathrm{circ}}(r_{\mathrm{ISCO}})$ in the Kerr-MOG spacetime.
    (a)~$\eta$ as a function of spin $a_*$ for $\alpha = 0,\,0.5,\,1.0,\,1.5$.
    The solid black curve is the standard Kerr black hole result. Increasing $\alpha$ progressively suppresses the efficiency, especially at high spin.
    (b)~$\eta$ as a function of $\alpha$ at fixed spins $a_* = 0,\,0.5,\,0.9,\,0.998$.
    The efficiency decreases monotonically with $\alpha$, with the steepest decline for near-extremal spin.}
  \label{fig:efficiency}
\end{figure*}

Figure~\ref{fig:efficiency}(a) shows $\eta$ as a function of spin $a_*$ for $\alpha = 0,\,0.5,\,1.0,\,1.5$. At $\alpha = 0$, the standard Kerr black hole result is recovered. $\eta$ rises steeply from $\eta \simeq 0.057$ at $a_* = 0$ to $\eta \simeq 0.32$ at $a_* = 0.998$. For $\alpha > 0$, the efficiency curves flatten markedly. At $\alpha = 1.0$, the maximum efficiency at $a_* = 0.998$ is only $\eta \simeq 0.11$, roughly a factor of three lower than the Kerr black hole value. Figure~\ref{fig:efficiency}(b) displays $\eta$ as a function of $\alpha$ at fixed spin. The suppression is most dramatic for rapidly spinning cases, where the ISCO is closest to the horizon in the $\alpha = 0$ limit: $\eta$ drops from $0.32$ to $\sim\!0.11$ as $\alpha$ increases from $0$ to $\sim\!1$ for $a_* = 0.998$. For slowly spinning black holes, the efficiency changes little because the Schwarzschild ISCO is already far from the horizon, and additional outward migration has a modest effect on binding energy.

\subsection{Photon Propagation}\label{sec:propagation}

Photons emitted from the disk propagate along null geodesics in the Kerr--MOG spacetime. Owing to the formal similarity with the Kerr--Newman spacetime, the geodesic equations can be recast into a Kerr--Newman form through the rescaling
\bqn
  \tilde{a} = \frac{a}{M_{\mathrm{eff}}}\,,\qquad
  \tilde{e} = \sqrt{\frac{\alpha}{1+\alpha}}\,,\qquad
  \tilde{r} = \frac{r}{M_{\mathrm{eff}}}\,,
\eqn
under which the horizon function becomes
$\tilde{\Delta} = \tilde{r}^{2} - 2\tilde{r} + \tilde{a}^{2} + \tilde{e}^{2}$,
the standard Kerr--Newman form with unit mass.
This allows efficient semi-analytic ray-tracing calculations via the YNOGK library \citep{Yang2013}.

Along each trajectory, the photon energy is modified by gravitational redshift and Doppler boosting. The total energy shift is described by the redshift factor
\bqn
  g \equiv \frac{E_{\mathrm{obs}}}{E_{\mathrm{em}}}
    = \frac{\sqrt{-g_{tt} - 2\Omega_K g_{t\varphi}
                 - \Omega_K^{2}\,g_{\varphi\varphi}}}
           {1 + \Omega_K\,\xi\,\sin i},
\eqn
where $i$ is the observer inclination angle (i.e., the angle between the disk normal and the line of sight) and $\xi \equiv -k_\varphi/k_t$ is the conserved azimuthal photon impact parameter. The redshift factor, therefore, depends on both the spacetime geometry and the photon trajectory.

\begin{figure*}[t]
  \centering
    \includegraphics[width=\textwidth]{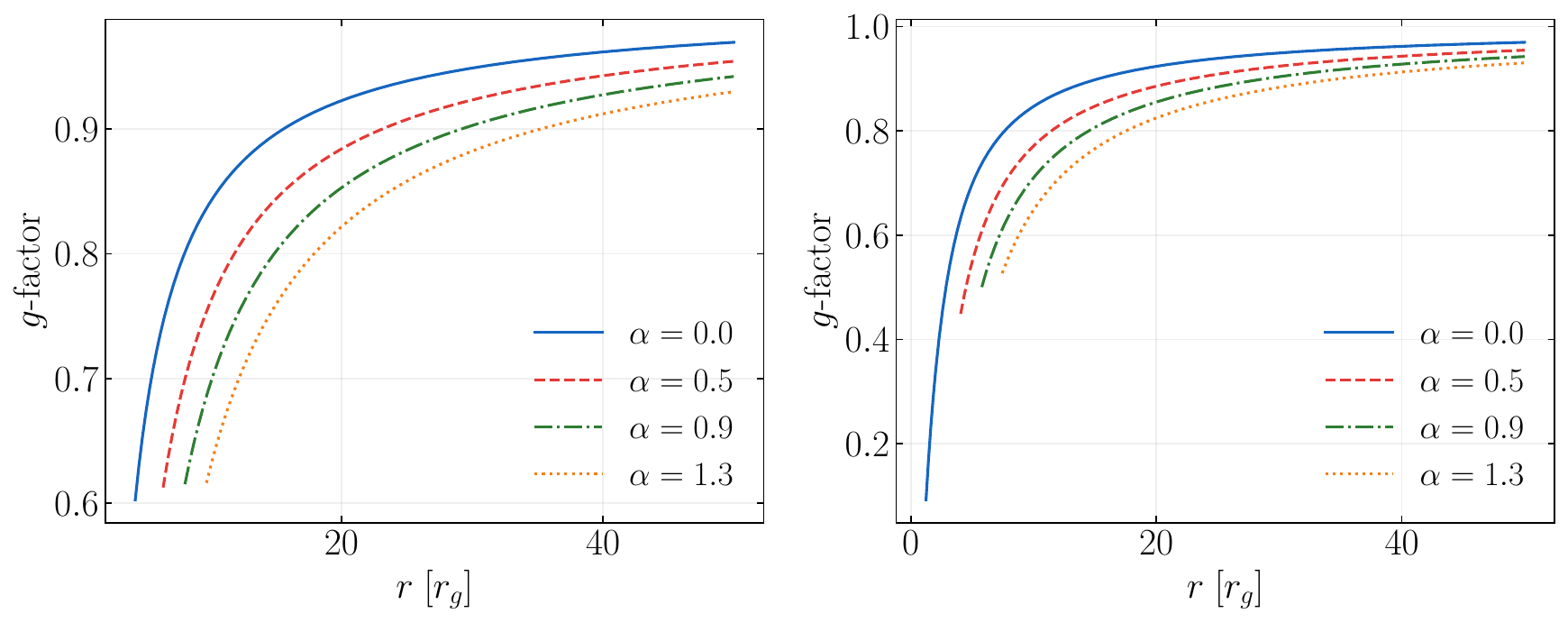}%
  \caption{%
    Gravitational redshift factor in the face-on limit ($i = 0$),
    $g\rvert_{i=0} = \sqrt{-g_{tt} - 2\Omega_K g_{t\varphi} - \Omega_K^2 g_{\varphi\varphi}}$,
    as a function of emitter radius for
    $\alpha = 0,\,0.5,\,0.9,\,1.3$ at
    $a_* = 0.5$ (left) and $a_* = 0.998$ (right). This quantity is the purely gravitational component of the full redshift factor $g$ defined in the text; the Doppler contribution from the photon trajectory ($1 + \Omega_K \xi \sin i$ in the denominator) vanishes for a face-on disk.}
  \label{fig:gfactor}
\end{figure*}

The metric components $g_{tt}$, $g_{t\varphi}$, and $g_{\varphi\varphi}$ at the equatorial plane carry explicit $\alpha$ dependence, thus, the combined gravitational redshift and Doppler boost differ from the Kerr black hole prediction even at fixed radius and orbital velocity. In practice, the approaching limb of the disk is slightly less blue-shifted in Kerr-MOG spacetime than in Kerr spacetime at the same spin, while the receding limb experiences a modest redshift. This asymmetry imprints itself on the high-energy wing of the thermal continuum, mildly steepening the spectral roll-off above the Wien peak.  Figure~\ref{fig:gfactor} shows the gravitational redshift factor $g(r)$ for emitters on circular orbits as a function of radius, comparing $\alpha = 0,\,0.5,\,0.9,\,1.3$ at two representative spins ($a_* = 0.5$ and $0.998$).

\subsection{Observed Spectrum}\label{sec:spectrum}

The specific intensity of the locally emitted radiation is modeled as a color-corrected blackbody (modified Planck function):
\bqn
  I_E^{\mathrm{em}}(E_{\mathrm{em}},\, r)
    = f_{\mathrm{col}}^{-4}\,
      B_E\!\left(E_{\mathrm{em}},\,f_{\mathrm{col}}\,T\right),
\eqn
where
\bqn
  B_E(E,\,T) = \frac{2E^{3}}{h^{2}c^{2}}
    \left[\exp\!\left(\frac{E}{kT}\right) - 1\right]^{-1}
\eqn
is the Planck function.
The color correction (or spectral hardening) factor $f_{\mathrm{col}}$ accounts for the fact that X-ray photons produced near the disk midplane may undergo predominantly electron scattering rather than true absorption on their way through the disk atmosphere.  If scattering dominates, the emergent photons preserve their higher midplane energies, and the observed spectrum appears harder than a pure blackbody at the effective temperature $T$. The color correction factor parameterizes this departure: photons are emitted as a blackbody at the hardened temperature $f_{\mathrm{col}}\,T$, while the prefactor $f_{\mathrm{col}}^{-4}$ ensures that the bolometric luminosity per unit area remains $\sigma T^{4}$, conserving the total radiated energy. In this work, we adopt the piece-wise prescription of \citet{Done2012}, in which $f_{\mathrm{col}}$ depends on the local temperature in three regimes: $f_{\mathrm{col}} \simeq 1$ for $kT \lesssim 2.6$~eV (local thermodynamic equilibrium regime), increases as $(kT/2.6\;\mathrm{eV})^{0.83}$ in an intermediate range, and saturates near $f_{\mathrm{col}} \simeq (72\;\mathrm{eV}/kT)^{1/9}$ at higher temperatures where electron scattering is dominant.

For each photon-ray traced from the disk to the observer, the relativistically invariant quantity $I_E / E^{3}$ is conserved along null geodesics. The observed specific flux at photon energy $E_{\mathrm{obs}}$ is therefore obtained by integrating over the image plane of the observer at distance $D$:
\bqn
  F_E(E_{\mathrm{obs}}) = \frac{1}{D^{2}}
    \int\!\!\int g^{3}\,
    f_{\mathrm{col}}^{-4}\,
    B_E\!\left(\frac{E_{\mathrm{obs}}}{g},\,f_{\mathrm{col}}\,T\right)
    db_\alpha\,db_\beta\,, \nb\\
\eqn
where $g = E_{\mathrm{obs}}/E_{\mathrm{em}}$ is the redshift factor defined in Section~\ref{sec:propagation}, and $(b_\alpha,b_\beta)$ are the photon impact parameters on the observer's sky. The factor $g^{3}$ arises from the Lorentz invariance of $I_E/E^{3}$: one power of $g$ converts the emitted energy to the observed energy, and two additional powers transform the solid angle element. The isotropic X-ray luminosity is then defined as
\bqn
  L_E \equiv 4\pi\,D^{2}\,F_E\,.
  \label{eq:LE}
\eqn
We plot $E L_E$ throughout this paper to display the spectral energy distribution.

In the \textsc{xspec} implementation, the observed photon number spectrum is computed as
\bqn
  \frac{dN}{dE}\bigg|_{\mathrm{obs}} = \frac{1}{D^{2}}
    \int\!\!\int g^{3}\,
    \frac{f_{\mathrm{col}}^{-4}\,B_E\!\left(E_{\mathrm{obs}}/g,\,f_{\mathrm{col}}\,T\right)}
         {E_{\mathrm{obs}}}\,
    db_\alpha\,db_\beta\,,  \nb\\
\eqn
and the distance enters through the model normalization parameter $K = 1/D_{\mathrm{kpc}}^{2}$, where $D_{\mathrm{kpc}}$ is the source distance in kiloparsecs.

This formulation self-consistently combines disk structure, radiative transfer through the color correction, relativistic photon transport, and observational projection into a unified framework for computing the thermal continuum spectrum in the Kerr--MOG spacetime.

\section{Thermal Continuum Spectra}
\label{sec:results}

\begin{figure*}[t]
  \centering
    \includegraphics[width=\textwidth]{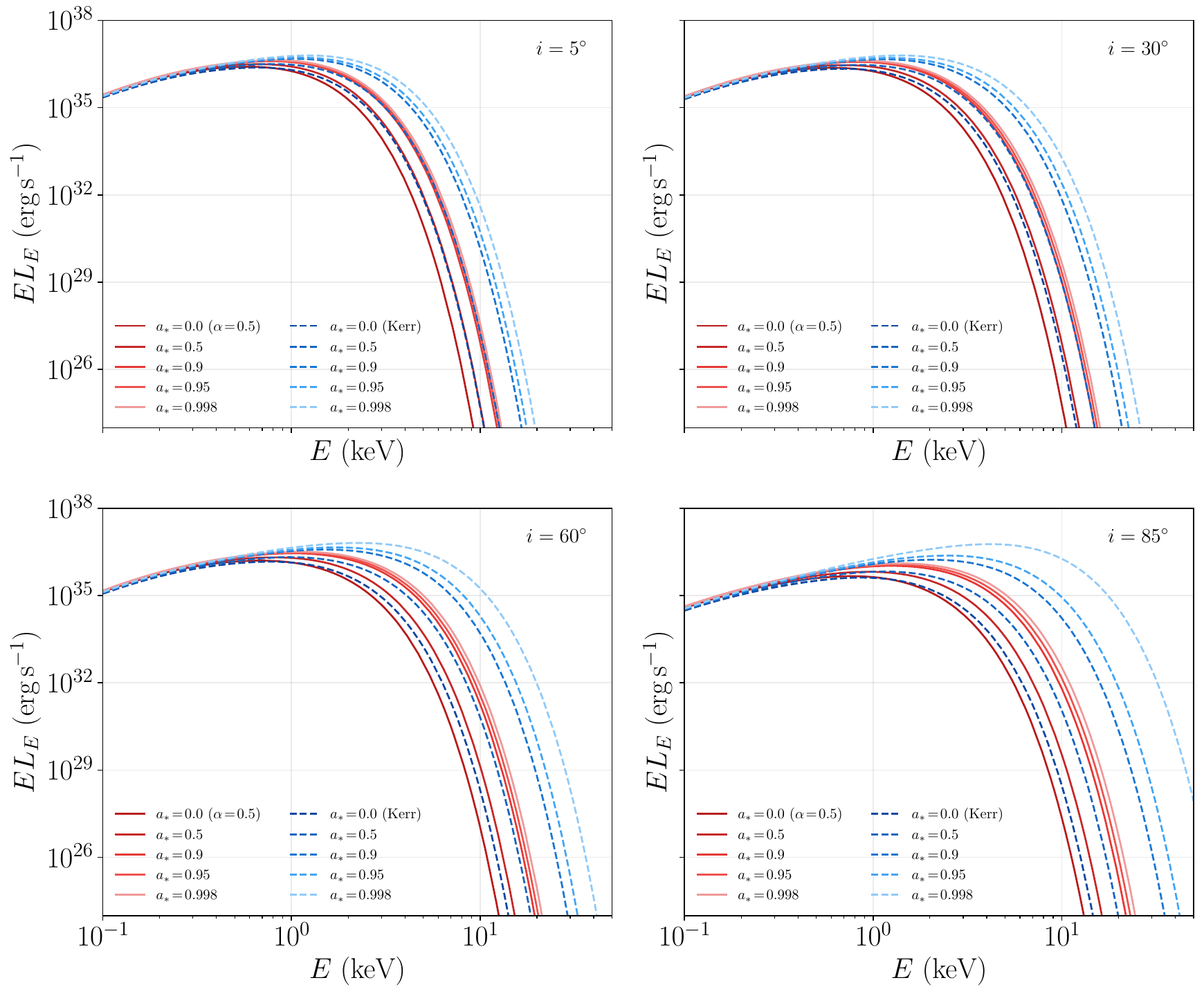}%
  \caption{%
    Thermal continuum spectra $EL_E$ computed by \texttt{kmspec} for a $10\,M_\odot$ black hole. Dashed blue: $\alpha = 0$ (Kerr limit); solid red: $\alpha = 0.5$ (Kerr-MOG). Each panel corresponds to a different observer inclination. Curves from bottom to top (at peak) correspond to different black hole spin $a_* = 0,\,0.5,\,0.9,\,0.95,\,0.998$.}
  \label{fig:4panel}
\end{figure*}

Using the formalism above, we first verified that all modified functions reproduce the standard Kerr spacetime results when $\alpha = 0$.
Figure~\ref{fig:4panel} shows the thermal continuum spectra computed by \texttt{kmspec} for $\alpha = 0$ (Kerr black hole, dashed blue curves) and $\alpha = 0.5$ (Kerr-MOG black hole, solid red curves), for spins $a_* \in \{0,\,0.5,\,0.9,\,0.95,\,0.998\}$ and inclinations $i \in \{5^\circ,\,30^\circ,\,60^\circ,\,85^\circ\}$. System parameters are $M = 10\,M_\odot$, $\dot{M} = 0.714\,\dot{M}_{\mathrm{Edd}}$, and $\delta_{\mathcal{J}} = 0$.

The Kerr-MOG black hole spectra are systematically softer (lower peak energy) and fainter than their Kerr black hole counterparts at the same spin. 
This is a direct consequence of the larger effective gravitational mass $M_{\mathrm{eff}} = 1 + \alpha$, which pushes the ISCO to larger radii and reduces the maximum disk temperature. 
The effect is qualitatively degenerate with a lower spin in the Kerr black hole,
highlighting the importance of independent spin measurements (e.g.,
from iron-line fitting) when constraining $\alpha$.

\begin{figure*}[t]
  \centering
    \includegraphics[width=\textwidth]{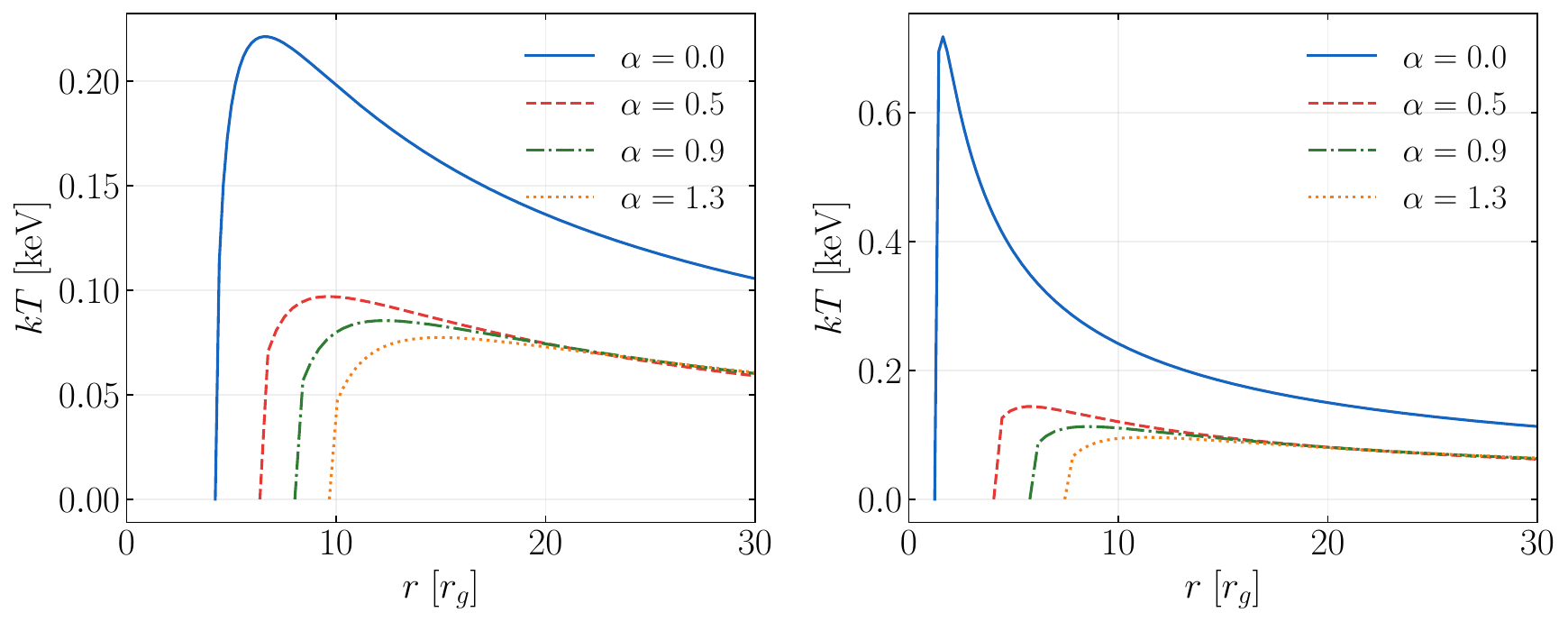}%
  \caption{%
    Radial temperature profiles $kT(r)$ of vanishing-ISCO-stress (Novikov--Thorne) thin disks for
    $\alpha = 0,\,0.5,\,0.9,\,1.3$.
    Left: $a_* = 0.5$; right: $a_* = 0.998$.
    Increasing $\alpha$ shifts the ISCO outward and lowers the peak temperature. System parameters:
    $M = 10\,M_\odot$, $\dot{M} = 0.1\,\dot{M}_{\mathrm{Edd}}$.}
  \label{fig:temperature}
\end{figure*}

The outward migration of the ISCO is the primary driver of the spectral softening: because the peak disk temperature scales approximately as $T_{\max} \propto r_{\mathrm{ISCO}}^{-3/4}$, a $20\%$ increase in the ISCO radius suppresses $T_{\max}$ by roughly $15\%$ and shifts the Wien peak to correspondingly lower energies.  Figure~\ref{fig:temperature} displays the radial temperature profile $kT(r)$ of the vanishing-ISCO-stress thin disk (Novikov--Thorne) for $\alpha = 0,\,0.5,\,0.9,\,1.3$ at two representative black hole spins ($a_*=0.5$ and $0.998$), illustrating how $\alpha$ modifies the thermal emission profile.

The magnitude of the fifth-force spectral signature is inclination-dependent. At low inclinations ($i \lesssim 15^\circ$), the Doppler effect is suppressed, and the dominant effect is the gravitational redshift, which depends on $\alpha$ through $g_{tt}$. At high inclinations ($i \gtrsim 70^\circ$), the Doppler effect is amplified and the altered orbital velocity $\Omega_K(\alpha)$ becomes the leading correction. Our model spectra (Figure~\ref{fig:4panel}) confirm that the fractional flux difference between $\alpha = 0$ and $\alpha = 0.5$ grows from $\sim\!5\%$ at $i = 5^\circ$ to $\sim\!20\%$ at $i = 85^\circ$ for a rapidly spinning black hole. It makes high inclination sources more promising targets for constraining
the vector-field strength.

\subsection{Physical interpretation and parameter degeneracies}
\label{sec:interp}

Perhaps the most consequential feature of the thermal spectrum in a Kerr-MOG black hole is its resemblance to a Kerr black hole spectrum of lower spin. Increasing $\alpha$ at fixed $a_*$ produces spectral changes, namely a cooler peak and reduced luminosity, that are nearly indistinguishable from decreasing $a_*$ at $\alpha = 0$, at least within the statistical precision of current X-ray missions. This degeneracy arises because the observed continuum is shaped primarily by the ISCO radius, and the mapping $r_{\mathrm{ISCO}}(a_*,\alpha)$ admits a family of $(a_*,\alpha)$ pairs that produce the same ISCO. In practice, a source well fitted by a Kerr black hole model at spin $a_*^{\mathrm{Kerr}}$ could equally be described by a Kerr-MOG black hole model with a higher intrinsic spin and a non-vanishing fifth-force charge.

Breaking this degeneracy requires either (i)~an independent spin measurement from a channel that responds differently to $\alpha$, such as the relativistically broadened iron K$\alpha$ line, whose profile is sensitive to both the ISCO and the emissivity index in a manner distinct from the continuum (\citep{Reynolds2014}) or (ii)~simultaneous fitting of multiple spectral components (continuum plus reflection) with a shared $(a_*,\alpha)$ pair, thereby exploiting the differential imprint of the fifth force on each component.

Several alternative theories of gravity predict deformed Kerr-like spacetimes that also modify the thermal continuum. For instance, Johannsen--Psaltis deformation parameters $\epsilon_n$ \citep{Johannsen2011} and the Konoplya--Rezzolla--Zhidenko parametrization \citep{Konoplya2016} introduce metric functions, whereas the Kerr-MOG metric has a definite physical origin in the STVG field equations.  The key practical distinction is that $\alpha$ simultaneously governs the ISCO location, the orbital kinematics, and the null geodesic structure through a single parameter, whereas generic deformation frameworks often allow these three aspects to vary independently. This enhanced predictive rigidity of the MOG model is both an advantage, in that it uses fewer free parameters, and a limitation, in that it offers less flexibility to accommodate anomalous data.

Recent analyses have demonstrated that super-extremal Kerr disks (i.e., with $a_* > 1$) can mimic non-Kerr signatures in the thermal continuum \citep{Mummery2024}. In the Kerr-MOG framework, the condition for the existence of an event horizon is $a^2 \le G_N^2(1+\alpha)^2 M^2$, which is less restrictive than the Kerr bound $a \le G_N M$ (equivalently $a_* \le 1$) for $\alpha > 0$. Thus, the fifth-force charge enlarges the domain of black-hole solutions that possess horizons, and some configurations that would be naked singularities in GR are regular black holes in STVG. This observation implies that joint constraints on $(a_*,\alpha)$ from thermal continuum fitting can distinguish STVG black holes from super-spinning compact objects postulated in other scenarios.

The spin--$\alpha$ degeneracy identified here carries an important corollary for the interpretation of existing black-hole spin surveys. If STVG is the correct theory of gravity, then all continuum-fitting spin measurements reported to date under the Kerr hypothesis would be systematically biased toward lower values: the true spin $a_*^{\mathrm{true}}$ of each source would satisfy $a_*^{\mathrm{true}} > a_*^{\mathrm{Kerr}}$, with the discrepancy growing with $\alpha$. Conversely, if independent spin measurements from reflection spectroscopy consistently exceed continuum-fitting values at high statistical significance, this tension could be interpreted as evidence for a non-zero $\alpha$. Current data offer no clear sign of such a systematic offset, since the spins of LMC~X-1, Cygnus~X-1, and GRS~1915+105 derived from continuum and reflection methods agree within their joint uncertainties \citep{Gou2009,Steiner2012,MillerJones2021}; nevertheless, the possibility warrants careful re-examination as both methods achieve higher precision with next-generation instruments.

The \texttt{kmspec} model is directly loadable into \textsc{xspec} and is therefore applicable to archival and forthcoming X-ray data from missions such as \textit{XRISM}, \textit{XMM-Newton}, \textit{NICER}, \textit{Insight-HXMT}, and the planned \textit{eXTP} and \textit{STROBE-X}. Although our framework is compatible with a wide range of observatories, we focus our subsequent performance tests specifically on the capabilities of \textit{XRISM} and \textit{XMM-Newton}. Ideal targets are persistent or recurrent X-ray binaries observed in the thermal-dominant state, where the disk emission overwhelms the Comptonized power-law tail. LMC~X-1, GRS~1915+105, and Cygnus~X-1 during soft-state excursions are prototypical candidates: each has a well-determined distance, dynamical mass measurement, and high-quality broadband spectra \cite[e.g.,][]{Podgorny:2023lze,Miller:2025aqd,Zhao:2021bhj,Steiner2010,Orosz2009}.

To test whether these theoretical signatures are practically measurable across different instrumental sensitivities, we now turn to instrument-level forward simulations focusing on \textit{XRISM}/Resolve and \textit{XMM-Newton}/EPIC-pn.

\section{Simulated X-ray Observations}\label{sec:xrism}

Following the physical interpretation above, we assess the detectability of the fifth-force signature with current-generation X-ray instruments by simulating observations of a geometrically thin accretion disk in Kerr-MOG spacetime with the Resolve micro-calorimeter spectrometer \citep{XRISMResolve2022} aboard the \textit{XRISM} \citep{Tashiro2020}. Resolve provides non-dispersive spectroscopy over the $0.3$--$12$~keV band with an energy resolution of $\Delta E \approx 5$~eV at 6~keV, making it a powerful instrument for resolving subtle spectral differences in thermal corona between Kerr and Kerr-MOG black holes.

\subsection{Simulation Setup}\label{sec:xrism_setup}

We generate synthetic XRISM/Resolve spectra using the \textsc{xspec} \texttt{fakeit} command with the Cycle-3 calibration files: the redistribution matrix file \texttt{rsl\_Hp\_L\_2025.rmf} (Hp-grade events, large pixel type), the ancillary response file \texttt{rsl\_pntsrc\_GVC\_2025.arf} (point source, gate-valve closed), and a nominal exposure time of $T_{\mathrm{exp}} = 100$~ks.
The spectral model is \texttt{TBabs*kmspec}, where \texttt{TBabs} accounts for interstellar photoelectric absorption with neutral hydrogen column density $N_H$.

As a representative astrophysical target, we adopt parameters motivated by the stellar-mass black hole Cygnus~X-1 \citep{MillerJones2021}: $M = 21.2\,M_\odot$, distance $D = 2.09$~kpc, and $N_H = 0.6 \times 10^{22}$~cm$^{-2}$. The model normalization is set to $K=1/D_{\mathrm{kpc}}^2 \approx 0.229$. We consider two scenarios, a Kerr black hole baseline and a mildly Kerr-MOG black hole case, summarized in Table~\ref{tab:xrism_scenarios}.

\begin{table}[htbp]
\caption{Parameters of the simulations. All cases adopt $M = 21.2\,M_\odot$, $\delta_{\mathcal{J}} = 0$, $N_H = 0.6 \times 10^{22}$~cm$^{-2}$, $D = 2.09$~kpc, and $T_{\mathrm{exp}} = 100$~ks.
\label{tab:xrism_scenarios}}
\begin{ruledtabular}
\begin{tabular}{lcccc}
Label & $a_*$ & $\alpha$ & $i\,(^\circ)$ & $\dot{M}/\dot{M}_{\mathrm{Edd}}$ \\
\hline
Kerr baseline   & 0.998 & 0.0 & 30 & 0.05 \\
MOG mild        & 0.998 & 0.5 & 30 & 0.05 \\
\end{tabular}
\end{ruledtabular}
\end{table}

Both cases share the same near-extremal black hole spin $a_* = 0.998$, moderate inclination $i = 30^\circ$, and accretion rate $\dot{M}/\dot{M}_{\mathrm{Edd}} = 0.05$, isolating the effect of the fifth-force strength on the thermal continuum while keeping all other parameters identical.

\subsection{Expected Spectral Differences}\label{sec:xrism_results}

\begin{figure}[t]
  \centering
    \includegraphics[width=\columnwidth]{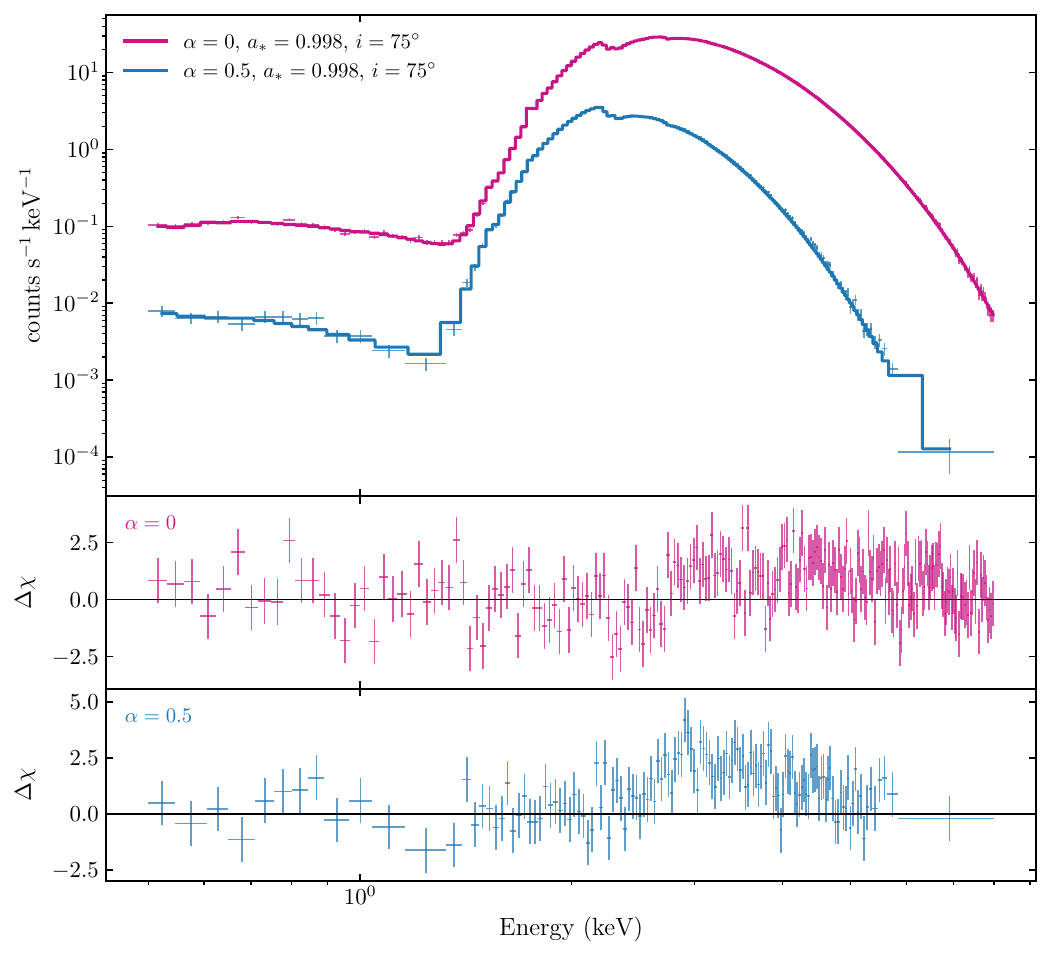}
  \caption{%
    Simulated XRISM/Resolve spectra comparing $\alpha = 0$ (magenta) and $\alpha = 0.5$ (blue), both at $a_* = 0.998$ and $i = 30^\circ$, each fitted with a pure Kerr black hole model.
    Upper panel: data (crosses) and best-fit model (histograms). Middle panel: $\Delta\chi$ residuals for the $\alpha = 0$ spectrum, showing no systematic structure.
    Lower panel: $\Delta\chi$ residuals for the $\alpha = 0.5$ spectrum, revealing the characteristic fifth-force signature.
    Cygnus~X-1 parameters ($M = 21.2\,M_\odot$, $D = 2.09$~kpc, $T_{\mathrm{exp}} = 100$~ks).}
  \label{fig:xrism_spectra}
\end{figure}

To quantify the detectability of the fifth-force signature, we simulate both Kerr black hole ($\alpha = 0$) and Kerr-MOG black hole ($\alpha = 0.5$) spectra with identical parameters ($a_* = 0.998$, $i = 30^\circ$, $\dot{M}/\dot{M}_{\mathrm{Edd}} = 0.05$), fit each with a pure Kerr black hole model (\texttt{TBabs*kmspec} with $\alpha = 0$ frozen, allowing $a_*$, $\dot{M}$, and normalization to vary), and compare the $\Delta\chi$ residuals side by side that are seen in Figures~\ref{fig:xrism_spectra} and~\ref{fig:xmm_spectra}.

For \textit{XRISM}/Resolve (Figure~\ref{fig:xrism_spectra}), the $\alpha = 0$ spectrum is well reproduced by its own model, with residuals randomly scattered about zero, confirming that the fitting procedure does not incur systematic bias. In contrast, the $\alpha = 0.5$ residuals display a coherent pattern: the best-fit Kerr black hole model over-predicts the flux near the Wien peak (where the ISCO-shifted temperature suppresses the MOG spectrum) and under-predicts at lower energies (where the cooler disk in the Kerr-MOG black hole contributes relatively more). 
The deviations per-bin reach $|\Delta\chi| \sim 3$--$5\sigma$, and the cumulative mismatch exceeds the detection threshold by a wide margin.

\subsection{Comparison with XMM-Newton/EPIC-pn}\label{sec:xmm}

We repeat the same paired analysis using the \textit{XMM-Newton} \citep{Jansen2001} EPIC-pn \citep{Struder2001} response. To ensure a controlled comparison between the two observatories, we employ the same underlying synthetic spectrum generated in Section~\ref{sec:xrism_results}. This model is folded through the epoch-4 full-frame redistribution matrix and ancillary response for EPIC-pn, assuming a 100~ks exposure and the previously defined Cygnus~X-1 parameters. By using an identical source model, we isolate the impact of instrumental resolution and effective area on the resulting MOG parameter constraints.

\begin{figure}[t]
  \centering
    \includegraphics[width=\columnwidth]{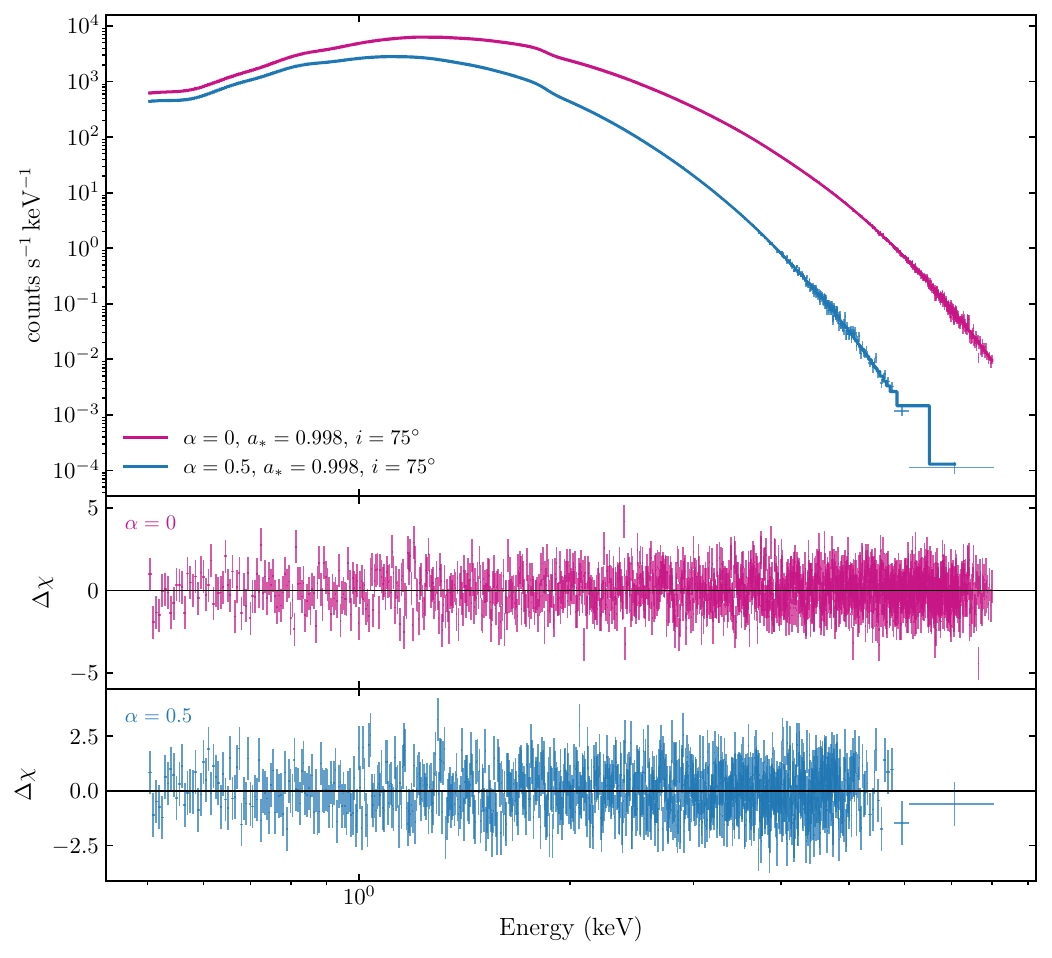}
  \caption{%
    Same as Figure~\ref{fig:xrism_spectra} but for simulated \textit{XMM-Newton}/EPIC-pn observations. The $\alpha = 0$ residuals (middle panel) are again consistent with zero, while the $\alpha = 0.5$ residuals (lower panel) reproduce the same systematic pattern. The broader energy resolution of EPIC-pn ($\sim\!150$~eV at 6~keV) smooths fine spectral features, but the substantially higher count rates ensure that the MOG residuals are detected at high significance.}
  \label{fig:xmm_spectra}
\end{figure}

Figure~\ref{fig:xmm_spectra} shows the resulting EPIC-pn spectral fits. Once again, the $\alpha = 0$ residuals are clean, while the $\alpha = 0.5$ residuals exhibit the same coherent pattern seen in Resolve, where they are positive at low energies and negative near the Wien peak. It is confirmed that the signature is an intrinsic property of the model rather than an instrument artifact. Because the EPIC-pn effective area ($\sim\!1200$~cm$^2$ at 1~keV) is roughly four times larger than that of Resolve, the count rate per resolution element is $\sim\!800$ times higher, and many more spectral bins contribute to the fit. Although the coarser energy resolution ($\sim\!150$~eV at 6~keV) smooths out fine spectral structure, the substantially higher count rates ensure that the broadband fifth-force signature is detected at very high statistical significance.

Conversely, the superb energy resolution of XRISM/Resolve ($\Delta E \approx 5$~eV at 6~keV) would provide a decisive advantage for constraining the MOG signature through narrow spectral features (e.g., the relativistically broadened iron K$\alpha$ line), where fine spectral structure carries the diagnostic information. 
For the smooth thermal continuum considered here, however, the higher photon statistics of EPIC-pn make it a more powerful broadband probe. In both instruments, the $\alpha = 0$ residuals act as a control, showing that the coherent structure at $\alpha = 0.5$ is genuinely driven by the fifth-force modification of the disk spectrum.

\begin{figure}[htb!]
  \centering
  \includegraphics[width=\columnwidth]{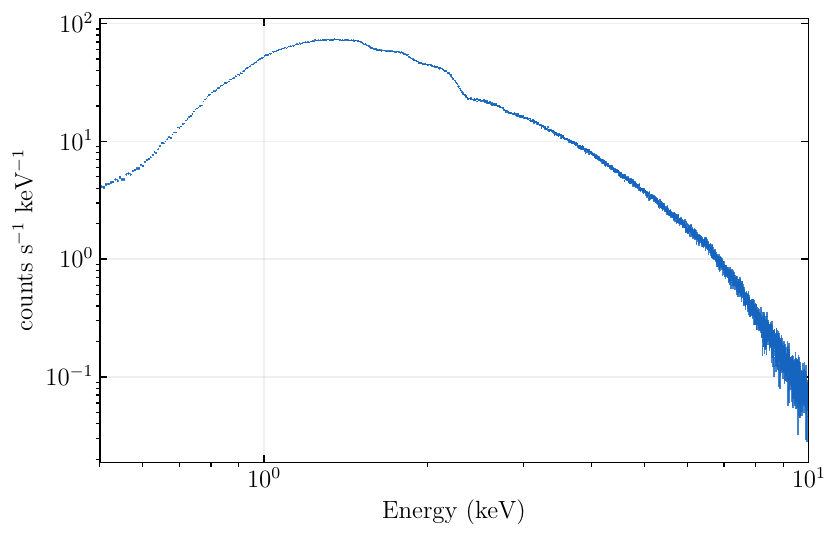}
  \caption{Background-subtracted \textit{XMM-Newton}/EPIC-pn count-rate spectrum of LMC~X-1 (ObsID~0743060101, 69.6~ks exposure) in the 0.5--10~keV band. The spectrum shows the characteristic thermal-dominant soft-state morphology, with the disk emission
  peaking at ${\sim}\,1$~keV.}
  \label{fig:xmm_real}
\end{figure}

\section{Constraints on $\alpha$ from Spectral Fitting}
\label{sec:constraints}

\subsection{XMM-Newton Observation of LMC~X-1}
\label{sec:xmm_obs}

\begin{figure*}[htb!]
  \centering
  \includegraphics[width=0.9\textwidth]{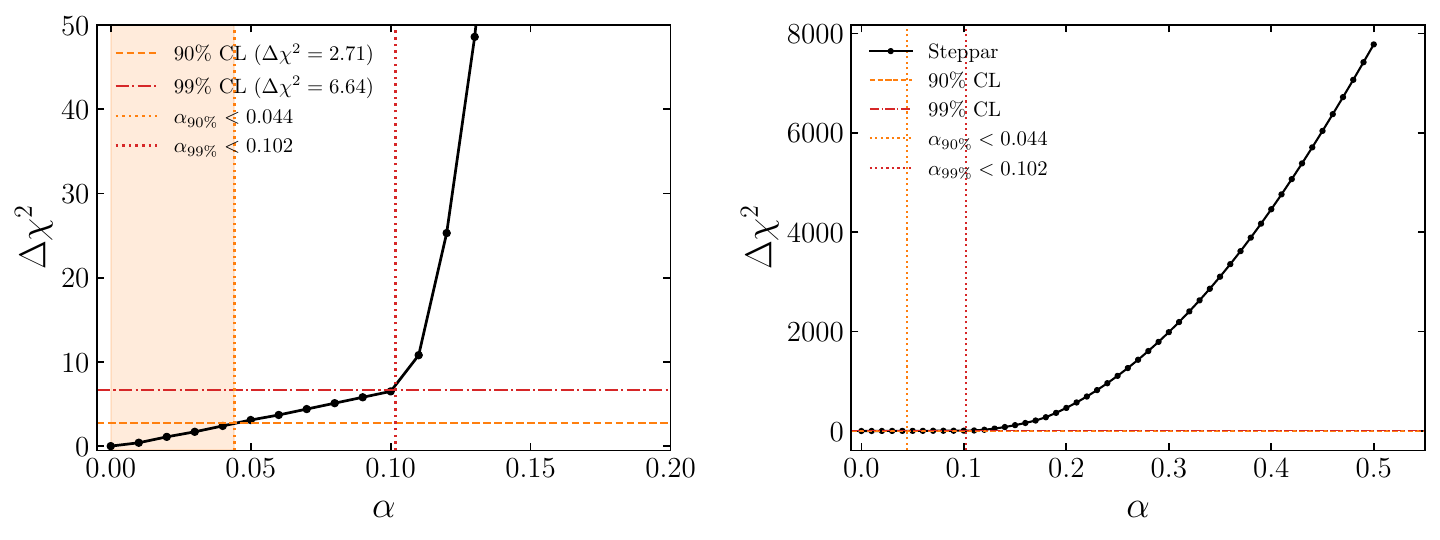}
  \caption{$\Delta\chi^2$ profile for $\alpha$ obtained from \texttt{steppar} applied to the \textit{XMM-Newton} EPIC-pn spectrum of LMC~X-1. The left panel shows the region near the minimum with the 90\% and 99\% confidence level thresholds. The right panel shows the full grid range $\alpha \in [0, 0.5]$. The steep rise demonstrates that even modest fifth-force couplings are strongly disfavored by the data.}
  \label{fig:chi2_real}
\end{figure*}

Building on the simulation-level detectability analysis above, we now fit real X-ray data to derive empirical bounds on $\alpha$.

LMC~X-1 is a persistent high-mass X-ray binary in the Large Magellanic Cloud that resides in the thermal-dominant state virtually at all epochs, making it an ideal target for continuum-fitting measurements of black hole spin \citep{McClintock2014}. Dynamical studies constrain the binary parameters to $M = 10.91 \pm 1.41\,M_\odot$, $i = 36.38^\circ \pm 1.92^\circ$, and $D = 48.1 \pm 2.2$~kpc \citep{Orosz2009}. Continuum fitting yields a spin $a_* = 0.92^{+0.05}_{-0.07}$ \citep{Gou2009}, in good agreement with the iron-line reflection measurement $a_* = 0.97^{+0.01}_{-0.13}$ \citep{Steiner2012}.

We analyze an archival \textit{XMM-Newton} observation \citep{Jansen2001} of LMC~X-1 (ObsID~0743060101, 2014 November~9) obtained with the EPIC-pn camera \citep{Struder2001} in full-frame mode with the thick optical-blocking filter. 
After standard screening, the net exposure is ${\approx}\,69.6$~ks with ${\sim}\,8.7 \times 10^6$ total source counts in the 0.5--10~keV band. A 2\% systematic uncertainty is added to all spectral bins to account for residual calibration effects.

Figure~\ref{fig:xmm_real} shows the background-subtracted EPIC-pn spectrum, which displays the characteristic thermal-dominant morphology: a prominent disk blackbody component peaking at ${\sim}\,1$~keV with a comparatively weak power-law tail extending to higher energies. The high count rate and soft spectral shape are fully consistent with previous observations of LMC~X-1 in its canonical soft state \citep{Gou2009}.

We fit the grouped spectrum with \texttt{TBabs*(kmspec + powerlaw)}, where the additive \texttt{powerlaw} accounts for the non-thermal high-energy tail. The inclination and black hole mass are frozen at the dynamically determined values ($i = 36.38^\circ$, $M = 10.91\,M_\odot$; \citealt{Orosz2009}), and the ISCO stress parameter is set to $\delta_{\mathcal{J}} = 0$. The seven free parameters are $N_H$, $a_*$, $\dot{M}/\dot{M}_{\mathrm{Edd}}$, $\alpha$, the distance normalization $K = 1/D_{\mathrm{kpc}}^2$, the photon index~$\Gamma$ (where $dN/dE \propto E^{-\Gamma}$), and the power-law normalization.

Levenberg--Marquardt minimization yields $\chi^2/\nu = 3186.1/1885 = 1.69$, with best-fit values $a_* = 0.92 \pm 0.16$, $\alpha = 5.3 \times 10^{-3} \pm 0.25$,
$N_H = 0.804 \times 10^{22}$~cm$^{-2}$, $\dot{M}/\dot{M}_{\mathrm{Edd}} = 1.94$, and $\Gamma = 3.17$. The formal super-Eddington value of $\dot{M}/\dot{M}_{\mathrm{Edd}}$ should be interpreted as a fit parameter in the simplified continuum plus power-law phenomenology rather than as a precise physical estimate of the thin-disk accretion-rate. The spin is in excellent agreement with the continuum-fitting measurement $a_* = 0.92^{+0.05}_{-0.07}$ of \citet{Gou2009} and the iron-line reflection result of \citet{Steiner2012}, although the fifth-force coupling is consistent with zero.

To obtain frequentist upper limits on the deformation parameter, we run \texttt{steppar} over $\alpha \in [0, 0.5]$ in 50 grid steps, re-optimizing all remaining parameters at each grid point. Figure~\ref{fig:chi2_real} shows the resulting $\Delta\chi^2$ profile. Because the best fit lies at $\alpha = 0$, the profile provides one-sided upper limits:
\begin{align}
  \alpha < \begin{cases}
    0.044 & (90\%\ \mathrm{CL}),\\
    0.062 & (95\%\ \mathrm{CL}),\\
    0.102 & (99\%\ \mathrm{CL}).
  \end{cases}
  \label{eq:alpha_UL_real}
\end{align}
These constraints indicate that the \textit{XMM-Newton} spectrum of LMC~X-1 is fully consistent with the Kerr metric and places stringent limits on the strength of any fifth-force modification of the ISCO structure.

\begin{figure*}[htb!]
  \centering
  \includegraphics[width=\textwidth]{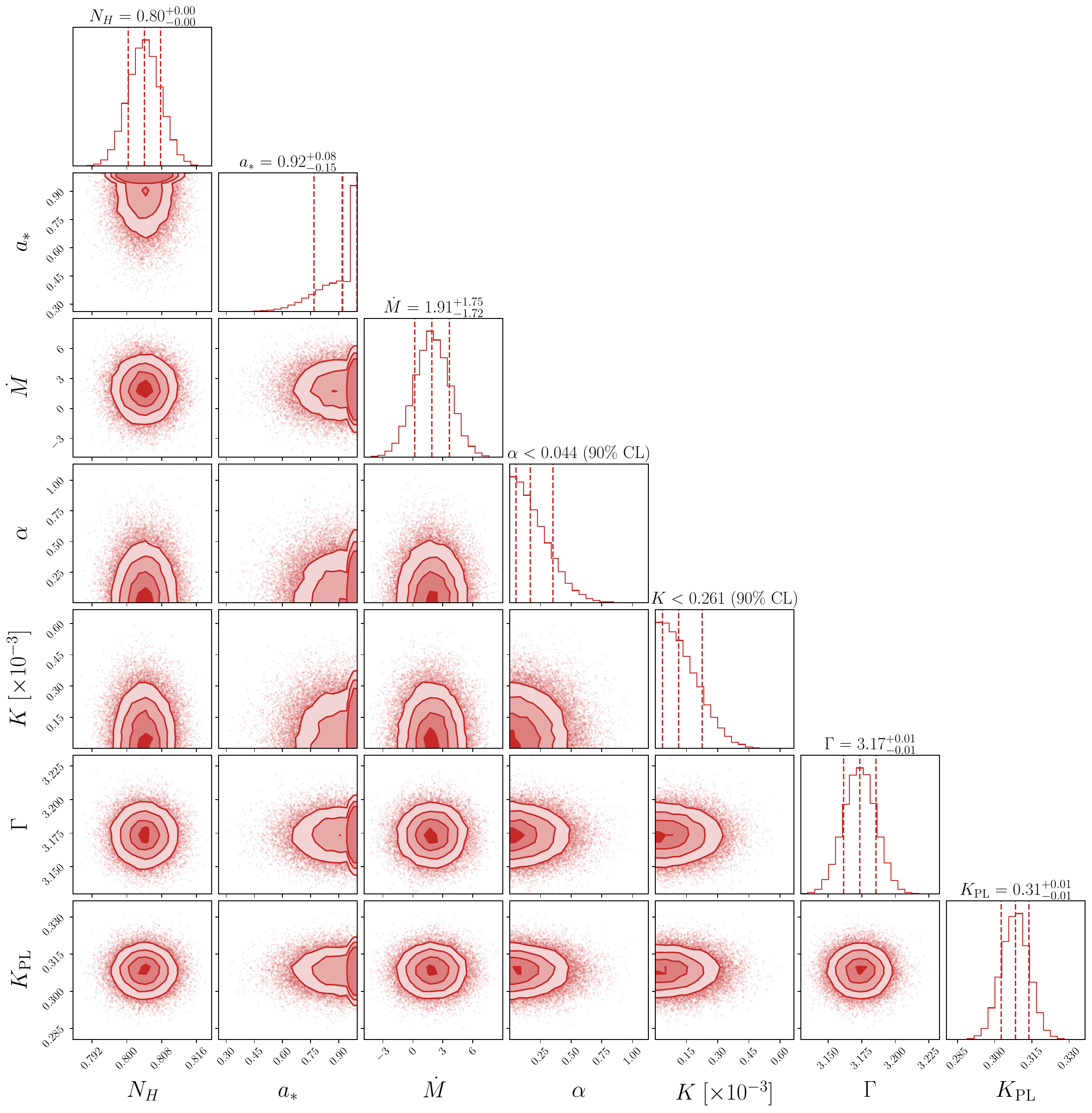}
  \caption{Corner plot of the marginalized posterior distributions for all seven free parameters of the \texttt{TBabs*(kmspec+powerlaw)} fit to the real \textit{XMM-Newton}/EPIC-pn spectrum of LMC~X-1, constructed from the Gaussian fit covariance matrix. 
  The vertical dashed lines mark the 16th, 50th, and 84th percentiles. The $\alpha$ posterior is sharply peaked near zero, confirming the $\Delta\chi^2$ constraints of Fig.~\ref{fig:chi2_real}.}
  \label{fig:corner_real}
\end{figure*}

To convert the frequentist profile into a Bayesian constraint, we construct the marginal posterior density $p(\alpha \mid \mathcal{D}) \propto \exp(-\Delta\chi^2/2)$, assuming a flat prior in $\alpha \ge 0$.
Figure~\ref{fig:corner_real} shows the corner plot of all seven free parameters, generated from the fit covariance matrix.
The marginal peaks of $\alpha$ are near zero which yield one-sided upper limits:
\begin{align}
  \alpha^{\mathrm{(post)}} < \begin{cases}
    0.062 & (90\%\ \mathrm{CL}),\\
    0.098 & (99\%\ \mathrm{CL}),
  \end{cases}
  \label{eq:alpha_UL_post}
\end{align}
in good agreement with the $\Delta\chi^2$ thresholds of Eq.~\eqref{eq:alpha_UL_real}. The posterior median is $\alpha = 0.020$, confirming that the data require no deviation from GR.

The physical interpretation of the limit $\alpha < 0.044$ requires careful consideration of the phenomenological behavior of $\alpha$ across different astrophysical scales. In the weak-field regime relevant to galactic dynamics, fits to galaxy rotation curves and Chandra X-ray cluster data yield $\alpha = 8.89 \pm 0.34$ \citep{Moffat:2013sja}, whereas binary pulsar timing measurements constrain the Yukawa coupling parameter to $\alpha = (2.40 \pm 0.02) \times 10^{-8}$ \citep{Deng:2009tg}. These differing values, which span over eight orders of magnitude, arise from different parameterizations in the weak-field and post-Newtonian regimes, and do not represent a universal ``running'' of a single coupling constant. In the strong-field regime, Solar System tests of planetary perihelion precession have ruled out the galaxy-scale MOG parameters at more than $3\sigma$ \citep{Iorio:2008sk}, while the theory possesses a density-dependent screening mechanism that suppresses deviations from GR in high-density environments \citep{Moffat:2014asa}. Taken together, these observations indicate that the effective fifth-force strength is environment-dependent. However, they do not provide a firm theoretical prediction for $\alpha$ at the $\sim\!10\,M_\odot$ scale probed by our X-ray spectral analysis. Our constraint $\alpha < 0.044$ therefore represents the first direct empirical bound on the Kerr-MOG deformation parameter at the stellar-mass black hole scale, filling a gap between the weak-field galactic regime and the compact-object regime. A decisive test of the environment-dependence would require applying \texttt{kmspec} to sources spanning several decades in mass, from stellar-mass X-ray binaries to active galactic nuclei.

\subsection{Prospects for Improved Constraints}

The real \textit{XMM-Newton} observation of LMC~X-1 provides $\alpha < 0.044$ at 90\% confidence (Eq.~\ref{eq:alpha_UL_real}), demonstrating that the high count rate of the archival observation (${\sim}\,8.7 \times 10^6$ source counts) tightly constrains the spectral shape and reduces the spin--fifth-force degeneracy. Nevertheless, the constraint remains limited by the large source distance ($D = 48.1$~kpc; \citealt{Orosz2009}).

Substantially tighter bounds, potentially at the $\mathcal{O}(10^{-3})$ level, can be expected from Galactic thermal-dominant X-ray binaries at smaller distances, such as Cygnus~X-1 during soft-state excursions ($D = 2.09$~kpc; \citealt{MillerJones2021}), LMC~X-3 ($D \approx 49$~kpc but with exceptionally stable soft-state morphology), or 4U~1957+11 ($D \approx 10$~kpc). Applications of the \texttt{kmspec} model to such targets represent the natural next step.

Beyond stellar-mass black holes, the \texttt{kmspec} model is equally applicable to supermassive black holes (SMBHs) in active galactic nuclei (AGN), offering a complementary probe of the Kerr-MOG deformation parameter at a fundamentally different mass scale. Although the theory does not provide a firm prediction for how $\alpha$ varies between the stellar-mass and SMBH regimes, obtaining independent measurements at both scales is essential for mapping any environment-dependence of the fifth-force coupling. The thermal emission from AGN accretion disks peaks in the ultraviolet, but the high-energy Wien tail extends into the soft X-ray band ($0.1\text{--}2$~keV), where current instruments such as \textit{XMM-Newton}/EPIC-pn and the future \textit{Athena}/X-IFU provide the sensitivity needed for continuum fitting. Joint analysis of the thermal soft excess and the relativistically broadened iron K$\alpha$ line in AGN, which would combine the continuum-based $\alpha$ sensitivity developed here with reflection-based spin diagnostics, would offer a powerful multi-scale probe of STVG across the black-hole mass spectrum.

\section{Conclusions}\label{sec:concl}

We have carried out a systematic investigation of how the MOG vector field, acting as a fifth force through a gravitational charge $Q \propto \sqrt{\alpha}\,M$, modifies the thermal continuum emission of geometrically thin accretion disks in the Kerr-MOG
spacetime. To this end, we developed \texttt{kmspec}, a relativistic spectral model available as a local \textsc{xspec} package, which self-consistently propagates the deformation parameter $\alpha$ through the ISCO determination, the Novikov--Thorne disk flux, the geodesic ray-tracing, and the gravitational energy shift. The model predicts a systematic outward shift of the ISCO, a lower peak temperature, and a softer thermal continuum relative to a Kerr black hole at the same spin, with the effect growing toward higher inclinations.

The thermal continuum alone is strongly degenerate with spin. Thus, independent constraints from reflection spectroscopy or joint multi-component fits remain essential for isolating $\alpha$. Nevertheless, the horizon condition for Kerr-MOG black hole is less restrictive than the Kerr bound, which means joint $(a_*,\alpha)$ fits can also
help distinguish STVG black holes from other compact-object scenarios.

Application of \texttt{kmspec} to a real 69.6~ks \textit{XMM-Newton} EPIC-pn observation of LMC~X-1 (ObsID~0743060101) constrains the fifth-force coupling to $\alpha < 0.044$ (90\% CL), consistent with the Kerr metric and GR.

Several approximations remain in the current implementation. We adopt the Novikov--Thorne no-torque condition at the ISCO, use an empirical color-correction prescription, and rely on the exact metric mapping to an effective Kerr--Newman form for ray-tracing. These assumptions are standard for continuum-fitting work, but a fully self-consistent treatment of Kerr-MOG disks will ultimately require dedicated general-relativistic magnetohydrodynamic (GRMHD) simulations and radiative-transfer modeling in the modified spacetime. The natural next step for future studies is to apply \texttt{kmspec} to closer Galactic soft-state binaries such as Cygnus~X-1 during thermal-dominant excursions \citep{MillerJones2021}, where the achievable constraints on $\alpha$ should be substantially tighter.

\begin{acknowledgements}
This research is supported by the National Key R\&D Program of China (Grant No.\,2023YFE0101200), the National Natural Science Foundation of China (Grant Nos.\,12273022 and 12511540053), and the Shanghai Municipality Orientation Program of Basic Research for International Scientists (Grant No.\,22JC1410600).
T.Z. is supported by the National Natural Science Foundation of China (Grants Nos.\,12275238 and 12542053), the Zhejiang Provincial Natural Science Foundation of China (Grants Nos.\,LR21A050001 and LY20A050002), the National Key R\&D Program of China (Grant No.\,2020YFC2201503), and the Fundamental Research Funds for the Provincial Universities of Zhejiang in China (Grant No.\,RF-A2019015). The simulations were performed on the TDLI-Astro cluster at Shanghai Jiao Tong University.
\end{acknowledgements}


\clearpage

\bibliography{sample701}{}
\bibliographystyle{aasjournalv7}

\end{document}